\magnification=\magstephalf
\tolerance10000
\baselineskip=0.1667truein plus 0.1pt minus 0.1pt

\font\msbm=msbm10
\font\bfvv=cmbx10 at 14.4truept

\vsize8truein

\hsize5truein

\parindent=0.125truein

\null
\vskip1.5truein
\newbox\naslov
\setbox\naslov=\hbox{\bfvv A} \vskip-\ht\naslov \leftline{{\bfvv
ON p-ADIC POWER SERIES}} \vskip2\baselineskip \noindent Branko
DRAGOVICH, Institute of Physics, P.O.Box 57, 11001 Belgrade,
Yugoslavia; \ \   dragovich@phy.bg.ac.yu \vskip4\baselineskip

\noindent
{\bf Abstract}
We obtained the region of convergence and the summation formula for some
modified generalized hypergeometric series (1.2). We also investigated
rationality of the sums of the power series (1.3). As a result
the series (1.4) cannot be the same rational number in all $\msbm\hbox{Z}_p$.
\vskip2\baselineskip


\vskip4\baselineskip

\noindent {\bf 1  \  Introduction } \vskip\baselineskip

\noindent We are interested in investigation of various properties
of some $p$-adic power series of the form
$$
\sum^\infty_{n=0} A_nx^n\ ,\eqno(1.1)
$$
where coefficients $A_n\in\msbm\hbox{Q}$ and variable
$x\in\msbm\hbox{Q}_p$. Such series are often encountered in
$p$-adic analysis [1] as well as in its applications in
mathematical and theoretical physics (for a review, see, e.g.
Refs. 2-5). Due to rationality of $A_n$, the series (1.1) can be
simultaneously considered in all $\msbm\hbox{Q}_p$ and in
$\msbm\hbox{R}$. It is of particular interest to find all rational
points for some classes of the series (1.1). Some previous
author's investigations on $p$-adic series of the form (1.1) were
presented at the Fourth International Conference on $p$-Adic
Analysis ([6] and references therein).

In this contribution we mainly consider some general properties of the series
$$
\sum^\infty_{n=0}a_nR_{k,l}(n)x^n\ ,\eqno(1.2)
$$
where $a_n$ are coefficients of the generalized hypergeometric series
and $R_{k,l}(n) = P_k(n)/Q_l(n)$ are rational functions in
$n\!\in\!\msbm\hbox{Z}_0 = \{0,1,2,\cdots\}$. We also examine in some
details the series
$$
\sum^\infty_{n=0}n!P_k(n)x^n\ ,\eqno(1.3)
$$
where $P_k(n)$ is a polynomial of degree $k$. In particular, we show that
$$
\sum^\infty_{n=0}n!\eqno(1.4)
$$
cannot be the same rational number in $\msbm\hbox{Z}_p$ for every $p$.

Note that in virtue of non-archimedean properties of $p$-adic
norm, the necessary condition is also the sufficient one for the
series (1.1) to be convergent, {\it i.e.} (1.1) is $p$-adic
convergent for some $x$ iff
$$
\mid A_nx^n\mid_p\to0\ , \qquad n\to \infty\ .\eqno(1.5)
$$

It is worth mentioning that Schikhof's book [1] contains an
excellent introductory course to analysis of $p$-adic series and,
if necessary, can be used to better understand some of our
considerations.

\vskip4\baselineskip \noindent {\bf 2  \  Generalized
Hypergeometric Series } \vskip\baselineskip \noindent Let $P_k(n)$
be a polynomial
$$
P_k(n) = C_kn^k+C_{k-1}n^{k-1}+\cdots+C_0\ ,\quad
0\not=C_k,C_{k-1},\cdots,C_0\in\msbm\hbox{Q}\ ,\eqno(2.1)
$$
in $n\in\msbm\hbox{Z}_0$  of degree $k$. Let also
$Q_l(n)$ be another polynomial of degree $l$,
$$
Q_l(n) = D_ln^l+D_{l-1}n^{l-1}+\cdots+D_0\ ,\quad
0\not=D_l,D_{l-1},\cdots,D_0\in\msbm\hbox{Q}\ , \eqno(2.2)
$$
with restriction $  Q_l(n)\not=0$ for every $n\in\msbm\hbox{Z}_0$.

We will call the $R$-modified generalized hypergeometric series
$$
\eqalign{&{}_rF_s(\alpha_1,\alpha_2,\cdots,\alpha_r;\beta_1,\beta_2,\cdots,\beta_s;
R_{k,l};x)\cr
&= \sum_{n=0}^\infty
{(\alpha_1)_n(\alpha_2)_n\cdots(\alpha_r)_n\over
(\beta_1)_n(\beta_2)_n\cdots(\beta_s)_n}R_{k,l}(n){x^n\over n!}\ ,\cr}\eqno(2.3)
$$
where $R_{k,l}(n)$ is a rational function
$$
R_{k,l}(n) = {P_k(n)\over Q_l(n)}\eqno(2.4)
$$
with polynomials $P_k(n)$ and $Q_l(n)$ defined by (2.1) and (2.2),
respectively, and $(u)_0 = 1, \ (u)_n = u(u+1)\cdots (u+n-1)$ for
$n\ge1$. When $R_{k,l}(n)\equiv1$ one gets the standard definition of
the generalized hypergeometric series.
\vskip\baselineskip

\noindent {\bf Proposition 1}  The $R$-modified hypergeometric
series defined by (2.3), where
$\alpha_1,\alpha_2,\cdots,\alpha_r\in\msbm\hbox{Z}_+ =
\{1,2,3,\cdots\}$ and $\beta_1,\beta_2\cdots\beta_s\in
\msbm\hbox{Z}_+$, is $p$-adically convergent in the region
$$
\mid x\mid_p<p^{r-s-1\over p-1}\ .\eqno(2.5)
$$

\bigskip

\noindent
{\it Proof:}  Note that
$$
(u)_n = {(u+n-1)!\over(u-1)!}\eqno(2.6)
$$
if $u\in\msbm\hbox{Z}_+$. Then $p$-adic norm of the general term
in (2.3) can be written as
$$
\eqalign{&\bigg\vert{(\beta_1-1)!\cdots(\beta_s-1)!\over(\alpha_1-1)!
\cdots(\alpha_r-1)}\bigg\vert_p
\bigg\vert{(\alpha_1+n-1)!\cdots(\alpha_r+n-1)!\over(\beta_1+n-1)!\cdots
(\beta_s+n-1)!}\bigg\vert_p\cr
&\times\bigg\vert{P_k(n)\over Q_l(n)}
\bigg\vert_p\bigg\vert{x^n\over n!}\bigg\vert_p\ .\cr}\eqno(2.7)
$$
Recall that
$$
\mid m!\mid_p = p^{-{m-\sigma_{m}\over p-1}}\ ,\quad m\in\msbm\hbox{Z}_+\
,\eqno(2.8)
$$
where $\sigma_m$ is the sum of digits in the expansion of $m$
over the base $p$. Since the first factor does not depend on $n$ and
$\vert P_k(n)\vert_p/\vert Q_l(n)\vert_p$ is bounded it suffices to analyse
$$
\bigg\vert{{(\alpha_1+n-1)!\cdots(\alpha_r+n-1)!\over(\beta_1+n-1)!\cdots
(\beta_s+n-1)!}}\bigg\vert_p\bigg\vert{x^n\over
n!}\bigg\vert_p\ .\eqno(2.9)
$$
For large enough $n$ (2.9) behaves like
$$
\bigg(p^{-{r-s-1\over p-1}}\mid x\mid_p\bigg)^n\ ,\eqno(2.10)
$$
which tends to zero as $n\to\infty$ if
$$
p^{-{r-s-1\over p-1}}\mid x\mid_p<1\ ,\eqno(2.11)
$$
what just gives (2.5).
\vskip\baselineskip

Note that (2.5) does not depend on the values of the parameters
$\alpha_1,\alpha_2,\cdots,\alpha_r$ and
$\beta_1,\beta_2,\cdots,\beta_s$ but only on their multiplicity $r$ and
$s$. For the Gauss series
$$
{}_2F_1(\alpha,\beta;\gamma;x) =
\sum^\infty_{n=0}{(\alpha)_n(\beta)_n\over(\gamma)_nn!} x^n\eqno(2.12)
$$
one obtains $\mid x\mid_p<1$, like $\mid x\mid_\infty<1$ in the real case.

Let us now turn to finding the corresponding summation formula.
\vskip\baselineskip \noindent {\bf Proposition 2}  Let
${}_rF_s(\alpha_1,\alpha_2,\cdots,\alpha_r;\beta_1,\beta_2,\cdots,\beta_s;R_{k,l};x)$
be an $R$-modified generalized hypergeometric series defined by
(2.3) with the region of convergence given by (2.5). Then the
following summation formula
$$
\eqalign{&\sum_{n=0}^\infty
{(\alpha_1)_n(\alpha_2)_n\cdots(\alpha_r)_n\over
(\beta_1)_n(\beta_2)_n\cdots(\beta_s)_nn!}\bigg[
{(\alpha_1+n)(\alpha_2+n)\cdots(\alpha_r+n)\over(\beta_1+n)(\beta_2+n)\cdots
(\beta_s+n)(n+1)}\cr
&\times {A_\mu(n+1)\over B_\nu(n+1)}x-{A_\mu(n)\over
B_\nu(n)}\bigg]x^n = -{A_\mu(0)\over B_\nu(0)}\cr}  \eqno(2.13)
$$
is valid, where $A_\mu(n)$ and $B_\nu(n)$ are polynomials in
$n\in\msbm\hbox{Z}_0$ of the form (2.1) and (2.2), respectively.

\bigskip
\noindent
{\it Proof:}  The left hand side of (2.13) can be rewritten in the form
$$
\eqalign{&\sum_{n=1}^\infty
{(\alpha_1)_n(\alpha_2)_n\cdots(\alpha_r)_n\over
(\beta_1)_n(\beta_2)_n\cdots(\beta_s)_n}{A_\mu(n)\over
n!B_\nu(n)}x^n\cr
&-\sum_{n=0}^\infty
{(\alpha_1)_n(\alpha_2)_n\cdots(\alpha_r)_n\over
(\beta_1)_n(\beta_2)_n\cdots(\beta_s)_n}
{A_\mu(n)\over
n!B_\nu(n)}x^n\cr}
$$
which, by mutual cancellation of all terms except term for $n = 0$,
gives just $-{A_\mu(0)\over
B_\nu(0)}$.

\vskip\baselineskip
Although based on a simple derivation, (2.13) leads to the rather
non-trivial results. Notice that always when
$$
R_{k,l}(n) = {(\alpha_1+n)(\alpha_2+n)\cdots(\alpha_r+n)\over(\beta_1+n)
(\beta_2+n)\cdots
(\beta_s+n)}{t\over n+1}
{A_\mu(n+1)\over
B_\nu(n+1)}-{A_\mu(n)\over
B_\nu(n)}\ ,\eqno(2.14)
$$
where $A_\mu(n)$ and $B_\nu(n)$ are arbitrary polynomials defined like
(2.1) and (2.2), respectively, if $x= t$ we have the resulting rational
sum of (2.3), which does not depend on $x$ and is equal to
$-A_\mu(0)/B_\nu(0)$. Of course, the parameter $t$ and the argument $x$
belong to the region of convergence (2.5).

A generalized hypergeometric series is defined by its parameters,
$\alpha_1,\alpha_2,\cdots,\alpha_r$ and $\beta_1,\beta_2,\cdots,\beta_s$.
For a given generalized
hypergeometric series there are many possibilities to choose rational
functions $R_{k,l}(n)$ (2.14) with the corresponding rational sums
$-A_\mu(0)/B_\nu(0)$. Let us notice some characteristic cases with
$B_\nu(n)\equiv1$. $A_\mu(n)$ may contain any partial or complete
product of factors in the denominator:
$\beta_1+n-1,\beta_2+n-1,\cdots,\beta_s+n-1,n$. In the case when
$A_\mu(n)$ includes $n$ as a factor then $A_\mu(0) = 0$ and the sum of
the corresponding series (2.13) will be also equal to zero. An extreme
case is
$$
A_\mu(n) = (\beta_1+n-1)(\beta_2+n-1)\cdots(\beta_s+n-1)nB_{\nu}(n)
\eqno(2.15)
$$
that gives in (2.14) the polynomial
$$
P_k(n) = (\alpha_1+n)(\alpha_2+n)\cdots(\alpha_r+n)tA_\mu(n+1)-A_\mu(n)
\eqno(2.16)
$$
instead of a rational function $R_{k,l}(n)$. Thus we have
$$
\sum_{n=0}^\infty
{(\alpha_1)_n(\alpha_2)_n\cdots(\alpha_r)_n\over
(\beta_1)_n(\beta_2)_n\cdots(\beta_s)_n}P_k(n){t^n\over n!} = 0\eqno(2.17)
$$
if $P_k(n)$ has the form (2.16).
\vskip4\baselineskip

\noindent {\bf 3 \ Series  $\sum^\infty_{n=0}n!\, P_k(n) \, x^n$}
\vskip\baselineskip

\noindent
This series can be regarded as a simple example of the $R$-modified
generalized hypergeometric series, {\it i.e.}
$$
{}_2F_0(1,1;P_k;x) = \sum^\infty_{n=0}n!P_k(n)x^n\ .\eqno(3.1)
$$
Because of its relative simplicity the series (3.1) is suitable
for examination of various $p$-adic properties. Power series (3.1)
is divergent in the real case. From (2.5) it follows that its
$p$-adic region of convergence is $\mid x\mid_p<p^{1/(p-1)}$ and
it yields in $\msbm\hbox{Q}_p$:
$$
x\in\msbm\hbox{Z}_p = \{x\in\msbm\hbox{Q}_p:\mid x\mid_p\le1\}\ .\eqno(3.2)
$$
As a consequence of (3.2) we may take for $x$ any integer and the
series (3.1) will be $p$-adically convergent for every prime $p$.

The corresponding summation formula is
$$
\sum^\infty_{n=0}n![(n+1)A_{k-1}(n+1)x-A_{k-1}(n)]x^n = -A_{k-1}(0)\ .\eqno(3.3)
$$
For $x = 1$ it can be rewritten in the more suitable form
$$
\sum^\infty_{n=0}n!(n^k+u_k) = v_k\ ,\eqno(3.4)
$$
where $(n+1)A_{k-1}(n+1)-A_{k-1}(n) = n^k+u_k, \ v_k = -A_{k-1}(0)$.
One can easily see that $u_k = A_{k-1}(1)-A_{k-1}(0)$.

It is very useful to have expressions for finite (partial) sums of
(3.4). \vskip\baselineskip \noindent {\bf Proposition 3}  If
$$
S^{(k)}_n = \sum^{n-1}_{i=0}i!i^k\ ,\quad k\in\msbm\hbox{Z}_0\ ,\eqno(3.5)
$$
then
$$
S^{(k+1)}_n = - \delta_{0k}-kS_n^{(k)}-\sum^{k-1}_{l=0}\pmatrix{k+1\cr l\cr}
S^{(l)}_n+n!n^k\eqno(3.6)
$$
is a recurrent relation, where $\delta_{0k}$ is the Kronecker symbol
($\delta_{0k} = 1$ if $k = 0$ and $\delta_{0k} = 0$ if $k\not=0$).

\bigskip
\noindent
{\it Proof:}
$$
\eqalign{S^{(k)}_n &= \delta_{0k}+
\sum^{n-2}_{i=0}(i+1)!(i+1)^k =
\delta_{0k}+\sum^{n-1}_{i=0}i!(i+1)^{k+1}-n!n^k \cr
&= \delta_{0k}+\sum^{k+1}_{l=0}\pmatrix{k+1\cr l\cr}S^{(l)}_n-n!n^k\ .\cr}
$$
\vskip\baselineskip
Applying successively the recurrent relation (3.6) we obtain summation
formula of the form
$$
\sum^{n-1}_{i=0}i!(i^k+u_k) = v_k+n!A_{k-1}(n)\ ,\eqno(3.7)
$$
where $A_{k-1}(n)$ is a polynomial of degree $k-1$ in $n$ with integer
coefficients. As an illustration, here are the first four examples:
$$
\eqalign{&\hbox{(a)} \ \sum^{n-1}_{i=0}i!i = -1+n!\ ,\cr
&\hbox{(b)} \ \sum^{n-1}_{i=0}i!(i^2+1) = 1+n!(n-1)\ ,\cr
&\hbox{(c)} \ \sum^{n-1}_{i=0}i!(i^3-1) = 1+n!(n^2-2n-1)\ ,\cr
&\hbox{(c)} \ \sum^{n-1}_{i=0}i!(i^4-2) = -5+n!(n^3-3n^2+5)\ .\cr}\eqno(3.8)
$$

In a similar way to the Proposition 3 one can obtain recurrent relations
for $u_k$ and $v_k$:
$$
\eqalign{&u_{k+1} = -ku_k-\sum^{k-1}_{l=1}\pmatrix{k+1\cr l\cr}u_l+1\ ,
\quad u_1= 0, \ k\ge1\ ,\cr
&v_{k+1} = -kv_k-\sum^{k-1}_{l=1}\pmatrix{k+1\cr l\cr}v_l-\delta_{0k}\ ,\quad
k\ge0\ .\cr}\eqno(3.9)
$$

\topinsert
\noindent
{\bf Table 1} contains the first eleven values of $u_k$ and $v_k$.
\def\prored{&\vbox to 4pt{}&&&&&&&&&&&&&&&&&&&&&&&\cr}%
\vskip4pt
\leftline{
\vbox{\offinterlineskip\halign{\tabskip=4pt#\vrule&\hfil$#$\hfil&
\vrule#&\hfil$#$\hfil&\vrule#&\hfil$#$\hfil&\vrule#&\hfil$#$
\hfil&\vrule#&\hfil$#$\hfil&\vrule#&\hfil$#$\hfil&\vrule#&\hfil$#$
\hfil&\vrule#&\hfil$#$\hfil&\vrule#&\hfil$#$\hfil&\vrule#&\hfil$#$
\hfil&\vrule#&\hfil$#$\hfil&\vrule#&\hfil$#$\hfil&\vrule#\tabskip=0pt\cr
\noalign{\hrule}%
\prored
&k&&1&&2&&3&&4&&5&&6&&7&&8&&9&&10&&11&\cr
\prored
\noalign{\hrule}%
\prored
&u_k&&0&&1&&-1&&-2&&9&&-9&&-50&&267&&-413&&-2180&&17731&\cr
\prored
\noalign{\hrule}%
\prored
&v_k&&-1&&1&&1&&-5&&5&&21&&-105&&141&&777&&-5513&&13209&\cr
\prored
\noalign{\hrule}%
}}%
}%
\vskip2\baselineskip
\endinsert

It is worth noting that $i^k+u_k$ in (3.7) is a simplified form of
$P_k(i)$ which gives rational sum of (3.1) if $x = 1$. Such $P_k(i) =
i^k+u_k$ are suitable to obtain a general expression for the series
(3.1) with rational sum at $x = 1$. In fact, the generalized form of
(3.7) is
$$
\sum^{n-1}_{i=0}i!P_k(i) = V_k+n!B_{k-1}(n)\ ,\quad k\ge1\ ,\eqno(3.10)
$$
where $P_k(i) = \sum^k_{r=0}C_ri^r$ with $C_0 = \sum^k_{r=1}C_ru_r$,
$V_k = \sum^k_{r=1}C_rv_r$, $B_{k-1}(n) = \sum^k_{r=1}$ $C_rA_{r-1}(n)$ and
$C_1,C_2,\cdots,C_k\in\msbm\hbox{Q}$.

The above consideration performed for $x = 1$ can be extended to other
positive integers $x$ with some other values of $u_k$ and $v_k$.

Let us turn now to the sum of the power series (3.1) and
investigate some of its rationality problem at
$x\in\msbm\hbox{Z}_+$. It is useful to start with the simplest
case, {\it i.e.} $P_k(n)\equiv1$. \vskip\baselineskip \noindent
{\bf Theorem 1}  Let $x$ be a given positive integer. If $p$-adic
sum of the power series
$$
\sum^\infty_{n=0}n!x^n\eqno(3.11)
$$
is a rational number then it cannot be the same in $\msbm\hbox{Z}_p$
for every $p$.

\bigskip
\noindent
{\it Proof:}  Suppose there is such $x = t\in\msbm\hbox{Z}_+$ that there
exists $p$-adic  rational sum
$$
\sum^\infty_{n=0}n!t^n = {a(t)\over b(t)}\,\quad \quad a(t)\in\msbm\hbox{Z}\ , \ b(t)\in\msbm\hbox{Z}_+\ \eqno(3.12)
$$
the same for every p.
Let $S_n(t)$ be
$$
S_n(t) = \sum^{n-1}_{i=0}i!t^i\ .\eqno(3.13)
$$
Since $i!t^i<(n-1)!t^{n-1}$ when $0\le i\le n-2$ one has $S_n(t) = 0!+1!t+2!t^2
+\cdots+(n-2)!t^{n-2}+(n-1)!t^{n-1}<(n-1)!(n-1)t^{n-1}+(n-1)!t^{n-1} =
n!t^{n-1}\le n!t^n$. Thus we have inequality
$$
0<S_n<n!t^n\ ,\quad n>2, \ t\ge1\ .\eqno(3.14)
$$
For a fixed $b(t)\in\msbm\hbox{Z}_+$ one can write
$b(t)S_n(t) = b(t)[0!+1!t+2!t^2
+\cdots+(n-2)!t^{n-2}+(n-1)!t^{n-1}]<b(t)[(n-1)!t^{n-1}+(n-1)!t^{n-1}] =
2b(t)(n-1)!t^{n-1}< n!t^n$
if $2b(t)<nt$, {\it i.e.}
$$
0<b(t)S_n(t)<n!t^n\ ,\quad 2b(t)<nt\ .\eqno(3.15)
$$
According to (3.12) one has
$a(t) = b(t)S_n(t) +n!t^nb(t)[1+(n+1)t+\cdots]$. Due to our assumption,
$b(t)[1+(n+1)+...]$ must be the same rational integer in all $\msbm\hbox{Z}_p$
and we get congruence
$$
a(t)\equiv b(t)S_n(t)(\hbox{mod} \ n!t^n)\ ,\quad n\ge0, \ t\ge1\ .\eqno(3.16)
$$
The value of $a(t)$ belongs to the one of the following three
possibilities: $(i) \ a(t)>0$, $(ii) \ a(t)<0$ and $(iii) \ a(t) = 0$.
Consider each of these possibilities. According to (3.15) and (3.16)
for large enough $n$ we have:
$$
\eqalign{(i) \ \ \ &0<a(t)<n!t^n\ ,\cr
&0<b(t)S_n(t)<n!t^n\ ,\cr
&a(t)\equiv b(t)S_n(t)(\hbox{mod} \ n!t^n)\ ;\cr}
$$
$$
\eqalign{(ii) \ \ \ &-n!t^n<a(t)<0\ ,\cr
&0<b(t)S_n(t)<n!t^n\ ,\cr
&a(t)\equiv b(t)S_n(t)(\hbox{mod} \ n!t^n)\ ;\cr}
$$
$$
\eqalign{(iii) \ \ \ &a(t) = 0\ ,\cr
&0<b(t)S_n(t)<n!t^n\ ,\cr
&a(t)\equiv b(t)S_n(t)(\hbox{mod} \ n!t^n)\ .\cr}\eqno(3.17)
$$
Analysing the conditions in (3.17) we find the following candidates
for solution:
$$
\eqalign{(i) \ \ \ &a(t) = b(t)S_n(t)\ ,\cr
(ii) \ \ \ &a(t) = b(t)S_n(t)-n!t^n\cr}\eqno(3.18)
$$
and $(iii)$ without solution. Since $a(t)$ must be a fixed integer
we conclude that the solutions (3.18), which depend on $n$, are
impossible.
\vskip\baselineskip

As a particular case of the Theorem 1 we have that the sum of the
series (1.4) cannot be the same rational number in all $\msbm\hbox{Z}_p$.
Note  an earlier assertion (see [1], p.17) that
$\sum_{n=0}^\infty n!$ cannot be rational in $\msbm\hbox{Z}_n$ for
every n.

\vskip\baselineskip \noindent {\bf Theorem 2}  For fixed $k$ and
$x$ the sum of the power series
$$
\sum^\infty_{n=0}n!n^kx^n\ ,\quad k\in\msbm\hbox{Z}_0\ , \
x\in\msbm\hbox{Z}_+\setminus\{1\}\ ,\eqno(3.19)
$$
cannot be the same rational number in $\msbm\hbox{Z}_p$ for every $p$.

\bigskip
\noindent {\it Proof:}  When $k= 0$ it follows from Theorem 1.
Dividing (3.3) by $x$, for $k\ge1$ one has
$$
\sum^\infty_{n=0}n![n^k+u_k(x)]x^n = v_k(x)\ ,\quad
x\in\msbm\hbox{Z}_p\setminus\{0\}\ ,\eqno(3.20)
$$
as a generalization of (3.4). Analysing the system of linear
equations for coefficients of the polynomial $A_{k-1}(n)$, which
follows from
$$
(n+1)A_{k-1}(n+1)-{A_{k-1}(n)\over x} = n^k+u_k(x)\ ,\eqno(3.21)
$$
we conclude that $u_k(x)$ has the form
$$
u_k(x) = {-1+xF_{k-1}(x)\over x^k}\ ,\eqno(3.22)
$$
where $F_{k-1}(x)$ is a polynomial in $x$ of degree $k-1$ with integer
coefficients (for $k = 1, \cdots,4$ see the Table 2).
The series (3.19) might be the same rational number in all $\msbm\hbox{Z}_p$
for some $x\in\msbm\hbox{Z}_+$ iff
$$
-1+xF_{k-1}(x) = 0\ .\eqno(3.23)
$$
However $F_{k-1}(x)$ is a polynomial with integer coefficients and eq.
(3.23) has no solutions in $x\in\msbm\hbox{Z}_+\setminus\{1\}$.
\vskip\baselineskip

Among  the series of the form
$$
\sum^\infty_{n=0}n!n^k\ ,\quad k\in\msbm\hbox{Z}_+\ ,\eqno(3.24)
$$
it is easy to see (Table 1 and (3.8)) that
$$
\sum^\infty_{n=0}n!n = -1\eqno(3.25)
$$
in $\msbm\hbox{Z}_p$ for every $p$. According to the Table 1
the sum of the series
$$
\sum^\infty_{n=0}n!n^k\ ,\quad k = 2,3,\cdots, 11\ ,\eqno(3.26)
$$
cannot be the same rational number (for a fixed $k$) in all
$\msbm\hbox{Z}_p$. \vskip\baselineskip \noindent {\bf Proposition
4} The sum of the series
$$
\sum^\infty_{n=0}n!n^{q+1}\ ,\quad q = \hbox{any \ of \ prime \
numbers}\ ,\eqno(3.27)
$$
cannot be the same rational number (for a fixed $q$) in $\msbm\hbox{Z}_p$
for every prime $p$.

\bigskip
\noindent
{\it Proof:}  According to the recurrent relations (3.9) one has for any prime
number $q$ that $u_{q+1}\equiv1(\hbox{mod} \ q)$ and
$v_{q+1}\equiv1(\hbox{mod} \ q)$. Thus, $u_{q+1}\not=0$ and $v_{q+1}$
is a rational integer.
\vskip\baselineskip
\topinsert
\noindent
{\bf Table 2}  Expressions for $u_k(x)$ and $v_k(x)$ $(k = 1,\cdots,4)$
illustrate some of our conclusions.
\vskip4pt
\def\prored{&\vbox to 4pt{}&&&&&&&&&\cr}%
\leftline{
\vbox{\offinterlineskip\halign{\tabskip=4pt#\vrule&\hfil$#$\hfil&
\vrule#&\hfil$#$\hfil&\vrule#&\hfil$#$\hfil&\vrule#&\hfil$#$\hfil&\vrule#&
\hfil$#$\hfil&\vrule#\tabskip=0pt\cr
\noalign{\hrule}%
\prored
&k&&1&&2&&3&&4&\cr
\prored
\noalign{\hrule}%
\prored
&u_k(x)&&{-1+x\over x}&&{-1+3x-x^2\over
x^2}&&{-1+6x-7x^2+x^3\over x^3}&&{-1+10x-25x^2+15x^3-x^4\over x^4}&\cr
\prored
\noalign{\hrule}%
\prored
&v_k(x)&&-{1\over x}&&{-1+2x\over x^2}&&{-1+5x-3x^2\over x^3}&&
{-1+9x-17x^2+4x^3\over x^4}&\cr
\prored
\noalign{\hrule}%
}}%
}%
\vskip2\baselineskip
\endinsert

It is unlikely that $\sum^\infty_{n=0}n!n^k$ is a rational number
if $k\not=1$. Thus there is a sense to introduce the following
\vskip.5cm \noindent {\bf Conjecture}  The sum of the series
$$
\sum^\infty_{n=0}n!n^k\ ,\quad k\in\msbm\hbox{Z}_0
$$
is a rational number in all $\msbm\hbox{Z}_p$ iff $k = 1$. Or, in
the more general form, $p$-adic sum of the power series
$$
\sum^\infty_{n=0}n!n^kx^n\ ,\quad k\in\msbm\hbox{Z}_0\ , \ x\in\msbm\hbox{Z}_+
$$
is a rational number iff $k = x = 1$. \vskip4\baselineskip

\noindent {\bf 4 \ Concluding Remarks} \vskip\baselineskip
\noindent It is worth noting that the $p$-adic power series
$$
\sum^\infty_{j=0}(n+1)_jx^j\ ,\quad x\in\msbm\hbox{Z}_+
$$
cannot be a rational integer in any $\msbm\hbox{Z}_p$ as well as
the same rational number in all $\msbm\hbox{Z}_p$.
This follows from identity
$$
\sum^\infty_{n=0}n!x^n = S_n(x)+n!x^n\sum^\infty_{j=0}(n+1)_jx^j
$$
and the proof of the Theorem 1.

It is clear that the $p$-adic hypergeometric series (2.12)
satisfies the corresponding  hypergeometric differential equation,
i.e.
$$
x(1-x)w^{\prime\prime}+[\gamma -(\alpha +\beta +1)x]w^{\prime}
-\alpha\beta w=0 \ ,
$$
where $w={}_2F_1(\alpha,\beta;\gamma;x)$. Let us also notice that
the $p$-adic series
$$
F_{\nu}(x)=\sum^\infty_{n=0}n!x^{n+\nu} \ , \quad \nu\in\msbm\hbox
{Z}_{+} \ ,
$$
is a solution of the following differential equation
$$
({d^\nu \over dx^\nu} - {1 \over x^{2\nu}}) F_\nu (x) = f_\nu (x) \  ,
\quad x\in \msbm\hbox{Z}_p \setminus \{0\} \ ,
$$
where
$$
f_\nu (x) = -\sum^{\nu - 1}_{l=0} {l!\over x^{\nu - l}}.
$$
The series
$$
F(x) = \sum^\infty_{n=0} n!x^n
$$
may be regarded as an analytic solution of the differential equation
$$
x^2 F^{\prime\prime} (x) + (3x - 1)F^\prime (x) + F(x) = 0.
$$

Many of the above results, obtained for $x\in \msbm\hbox{Z}_{+}$,
may be extended to $x\in \msbm\hbox{Z}\setminus\{0\}$ and it will be
done elsewhere.

\vskip\baselineskip

\noindent {\bf Acknowledgments} The author wishes to thank the
organizers of the Fifth International Conference on $p$-Adic
Analysis for invitation and hospitality, Prof. L. Van Hamme for
discussions, and especially Prof. W. H. Schikhof for discussions
and some informal communications. \vskip4\baselineskip \noindent
{\bf References} {\baselineskip=2\baselineskip
\parindent=17pt

\item{[1]} W.H. Schikhof. Ultrametric Calculus - An Introduction
to $p$-Adic Analysis. Cambridge: Cambridge University Press, 1984.

\item{[2]} L. Brekke, P.G.O. Freund. $p$-Adic Numbers in Physics.
Phys. Rep.  233: 1-66, 1993.

\item{[3]} V.S. Vladimirov, I.V. Volovich, E.I. Zelenov. $p$-Adic
Analysis and Mathematical Physics. Singapore: World Scientific,
1994.

\item{[4]} A. Khrennikov. $p$-Adic Valued Distributions in
Mathematical Phy\-s\-ics. Dordrecht: Kluwer Academic Publishers,
1994.

\item{[5]} A. Khrennikov. Non-Archimedean Analysis: Quantum
Paradoxes, Dynamical Systems and Biological Models. Dordrecht:
Kluwer Academic Publishers, 1997.

\item{[6]} B. Dragovich. On Some $p$-Adic Series with Factorials.
In W.H. Schikhof, C. Perez-Garcia, J. Kakol eds. $p$-Adic
Functional Analysis. Lecture Notes in Pure and Applied
Mathematics. Vol. 192. New York: Marcel Dekker, 1997. pp 95-105;
math-ph/0402050.

\end